\documentclass[]{spie}

\usepackage{amsmath,amsfonts,amssymb}
\usepackage{graphicx}
\usepackage{wrapfig}
\usepackage[colorlinks=true, allcolors=blue]{hyperref}

\title{Liger for Next-Generation Keck Adaptive Optics:  Opto-Mechanical Dewar for Imaging Camera and Slicer}

\author[a,b]{James Wiley}
\author[a,b]{Kalp Mathur}
\author[b]{Aaron Brown}
\author[a,b]{Shelley A. Wright}
\author[a,b]{Maren Cosens}
\author[b]{Jerome Maire}
\author[c]{Michael Fitzgerald}
\author[d]{Tucker Jones}
\author[e]{Marc Kassis}
\author[c]{Evan Kress}
\author[f]{Renate Kupke}
\author[c]{James E. Larkin}
\author[e]{Jim Lyke}
\author[c]{Eric Wang}
\author[e]{Sherry Yeh}

\affil[a]{Department of Physics, University of California San Diego, USA}
\affil[b]{Center for Astrophysics and Space Sciences, University of California San Diego, USA}
\affil[c]{Department of Physics \& Astronomy, University of California Los Angeles, USA}
\affil[d]{Department of Physics, University of California Davis, USA}
\affil[e]{W.M. Keck Observatory, Waimea, HI}
\affil[f]{Department of Astronomy \& Astrophysics, University of California Santa Cruz, USA}

\authorinfo{Further author information: (Send correspondence to J.W.)\\J.W.: E-mail: jhwiley@ucsd.edu}

\begin{document} 
\maketitle

\begin{abstract}
Liger is a next generation adaptive optics (AO) fed integral field spectrograph (IFS) and imager for the W. M. Keck Observatory. This new instrument is being designed to take advantage of the upgraded AO system provided by Keck All-Sky Precision Adaptive-optics (KAPA). Liger will provide higher spectral resolving power (R$\sim$4,000-10,000), wider wavelength coverage ($\sim$0.8-2.4 $\mu$m), and larger fields of view than any current IFS. We present the design and analysis for a custom-made dewar chamber for characterizing the Liger opto-mechanical system. This dewar chamber is designed to test and assemble the Liger imaging camera and slicer IFS components while being adaptable for future experiments. The vacuum chamber will operate below $10^{-5}$ Torr with a cold shield that will be kept below 90 K. The dewar test chamber will be mounted to an optical vibration isolation platform and further isolated from the cryogenic and vacuum systems with bellows. The cold head and vacuums will be mounted to a custom cart that will also house the electronics and computer that interface with the experiment. This test chamber will provide an efficient means of calibrating and characterizing the Liger instrument and performing future experiments.
\end{abstract}

\keywords{Integral Field Spectrograph, Vacuum, Dewar, Calibration, Characterization, Infrared, Cryogenic, Adaptive Optics}

\section{INTRODUCTION}
\label{sec:intro}

\begin{figure}[!ht]
   \begin{center}
   \begin{tabular}{c}
   \includegraphics[width=1\textwidth,angle=0,trim={0in 0in 0in 0in},clip]{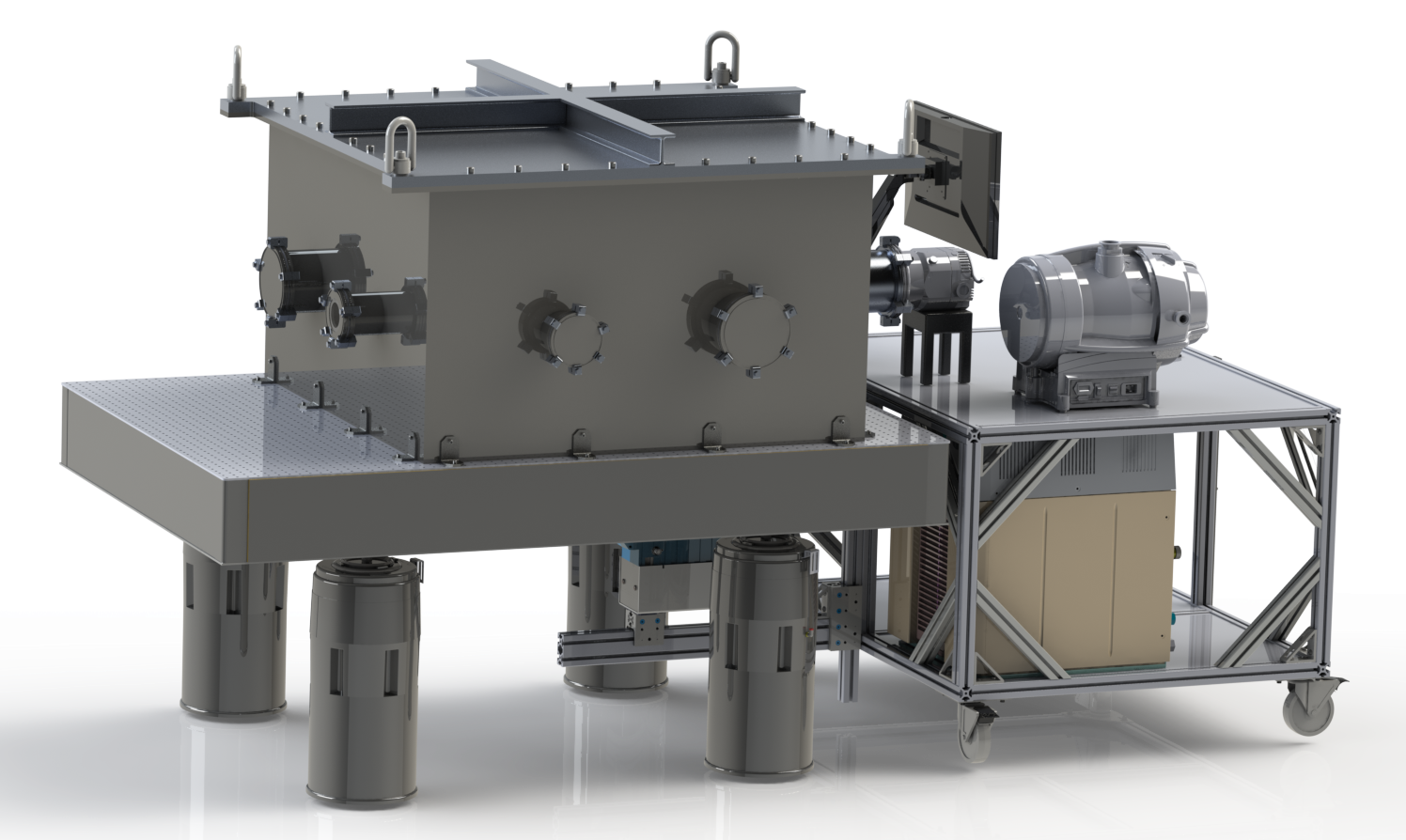}
   \end{tabular}
   \end{center}
   \caption[example]
  {
A rendering of the Liger test chamber experimental setup. The test chamber rests on a 4'$\times$6' optical table with active pneumatic vibration isolation supports.  The cold head that rises beneath the chamber and the vacuum connected to the side are housed on a cart made from 8020 components. The cold head and vacuums are vibrationally isolated from the chamber with bellows, and the cart is maneuverable and capable of mounting to the system from multiple different positions.}
\label{fig:1}
\end{figure}

Liger is a near-infrared ($\sim$0.8-2.4 $\mu$m) adaptive optics-fed imager and integral field spectrograph for the Keck I observatory \cite{Wright2020}.  This paper describes the design and analysis of the two cryogenic test chambers that will characterize the imager (at UCSD) and slicer (at UCSC) components of the Liger system. The Liger test chamber is a vibration isolated, cryogenically cooled, vacuum chamber for testing infrared optics and electronics (Fig. \ref{fig:1}).

The vacuum chamber is constructed from AISI 304 steel with a 6061 T6 aluminum lid and rests on a 4’x6’ optical table with pneumatic isolators. There is a flange on the bottom of the chamber and a through hole in the table to accommodate the cold head that extends into the chamber. The centered NW-160 port on one side of the chamber or an off-center NW-160 port on the sides are aligned with the cold head flange and can be used for the vacuum interface.  Both the cryogenic and vacuum systems are connected with bellows to reduce induced vibrations into the chamber. A NW-100 Flange with a custom built calcium fluoride window is used as the entrance for the light beam.

A 6061 T6 aluminum cold shield is connected to the base of the chamber with G-10 A-frames. Openings are cut in the cold shield for the optical beam entrance and the cold head, which extends into the cold shield and is thermally connected by copper strands. The 3 sides of the cold shield not facing the entrance beam have a 15cmx15cm slide-in feed through plate. There is an adjustable plate between the base of the cold shield and the optical platform to avoid induced flexure from thermal expansion and allow the system to be easily installed and aligned in the final dewar.

A cart made from 8020 holds the rough pump, turbo pump, cold head, compressor, and electronics that service the chamber.  The cold head is mounted on a cantilever arm that extends from the cart. It is connected to a two dimensional railing system to simplify positioning under the chamber. A 10" long bellows connects the cold head to the cold head flange. The vacuums are mounted on top of the cart close to the chamber so as to reduce the length of the vacuum line.  The compressor and electronics rack are held inside the cart.

The design of the experimental setup is based on previous characterization chambers \cite{Wiley_2017} \cite{2018_Fleming}. It will be used as a long term system for experiments after the characterization of the Liger imager and slicer. The experimental setup provides an accurate and precise way of characterizing infrared optics and electronics while being efficient and adaptable for future experiments.

\section{Vacuum Chamber}

\begin{figure}[ht]
\minipage{0.499\textwidth}
  \includegraphics[width=0.95\linewidth,trim={0.0in 0.0in 0.0in 0.0in}]{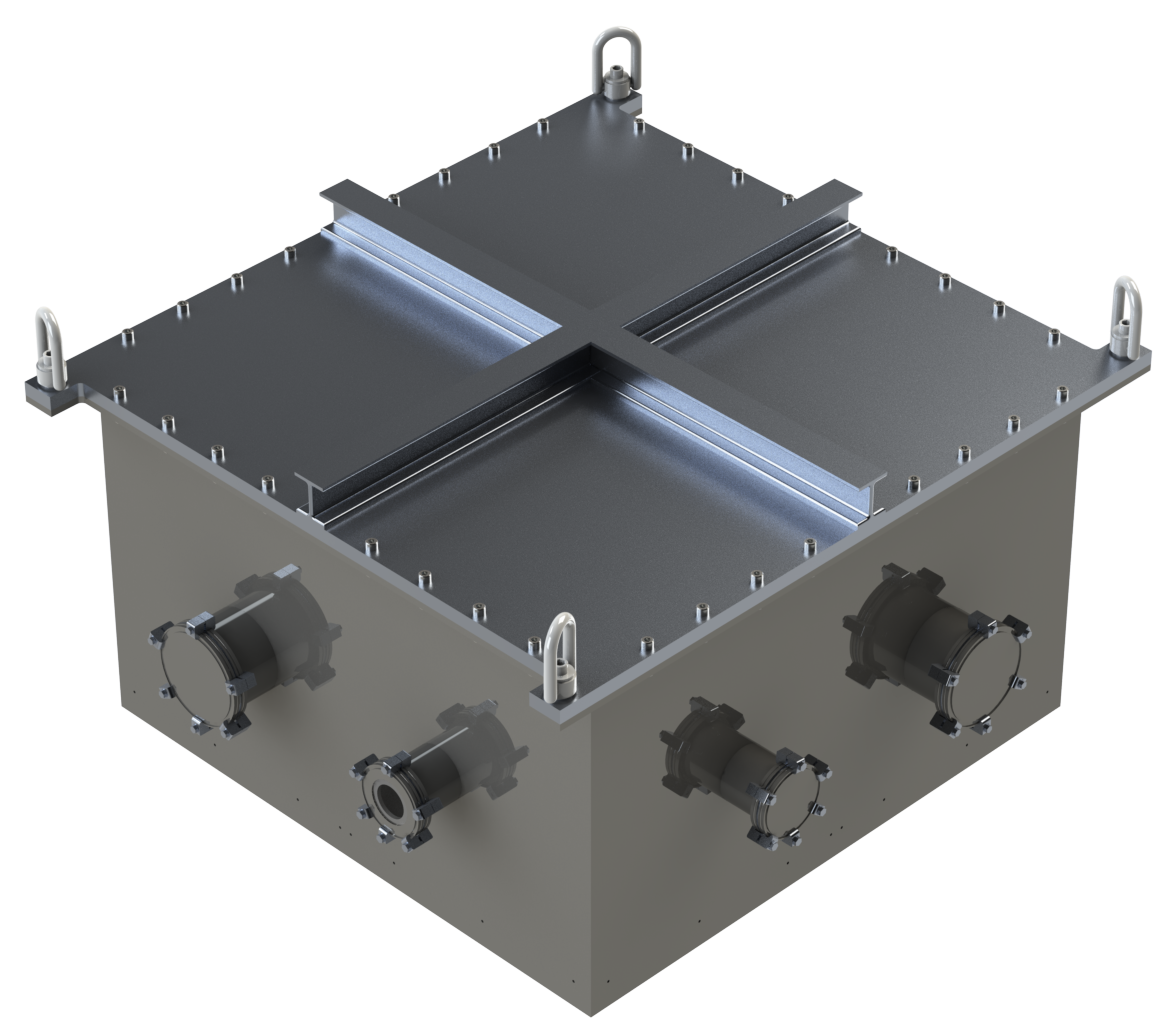}
\endminipage\hfill
\minipage{0.499\textwidth}
  \includegraphics[width=0.95\linewidth,trim={0.0in 0.0in 0.0in 0.0in}]{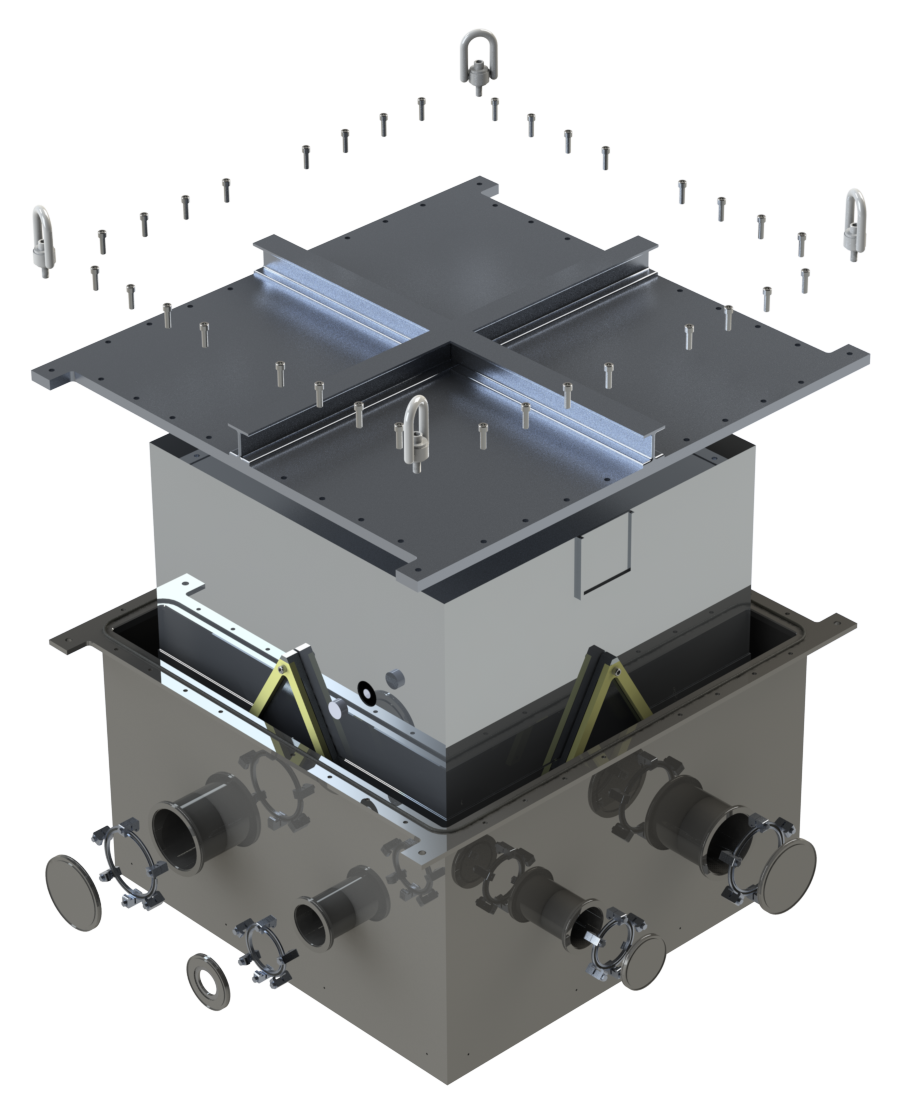}
\endminipage\hfill
\vspace{0.2in}
   \caption[example]
  {
Solid model and exploded view of the vacuum chamber showing the lid, entrance window, and cold shield that will be housed inside the chamber.}
\label{fig:2}
\end{figure}

The vacuum chamber rests on an 8" thick optical table and houses the cold shield and optics of the experimental setup (Fig. \ref{fig:2}).  It operates below a pressure of $\rm10^{-5}$ Torr. The body of the vacuum chamber is made of AISI 304 steel for its strength, low porosity, and favorable vacuum properties.  The lid for the chamber is made of 6061 T6 aluminum to decrease weight and has a cross of I-beams welded on for increased stiffness. 

The vacuum chamber inner dimensions are 1100mm$\times$1100mm$\times$623mm.  The total weight of the empty vacuum chamber is 1660 pounds while the weight of the lid alone is 250 pounds. There are three NW-100 flanges, four NW-160 flanges, and one cold head flange. Corner lips with mounting holes are located in the lid and top flange of the chamber for installing hoist rings. The rings thread into the lid and can be bolted onto the chamber with a nut for maneuvering just the lid or the entire chamber.  There are 1/4"-20 blind holes around the base of the chamber for fastening to a bracket mounted onto the optical table. This provides slight support against earthquakes and other such events.

The base of the chamber is 1” thick steel to withstand the force on it from the vacuum. The walls of the chamber are 5/8" thick and welded onto the base. The top flange is 1” wide and thick, and is welded on top of the chamber walls. The lid contains a dovetail O-ring groove and screws into the top flange to form the vacuum seal. The lid of the chamber is 1" thick 6061 T6 aluminum.  This material was chosen for its good vacuum properties, and it is lighter than steel making the lid easier to handle. A cross made of I beams is welded to the top of the lid which offers additional support against deflection.

A bolt hole pattern around the base of the inside of the chamber is for installing A-frames that connect to the cold shield.  A second set of bolt holes are centered on the base for installation of extra components and future experiments.

\subsection{Vacuum Analysis}

\begin{figure}[ht]
\minipage{0.499\textwidth}
  \includegraphics[width=0.95\linewidth,trim={0.0in 0.0in 0.0in 0.0in}]{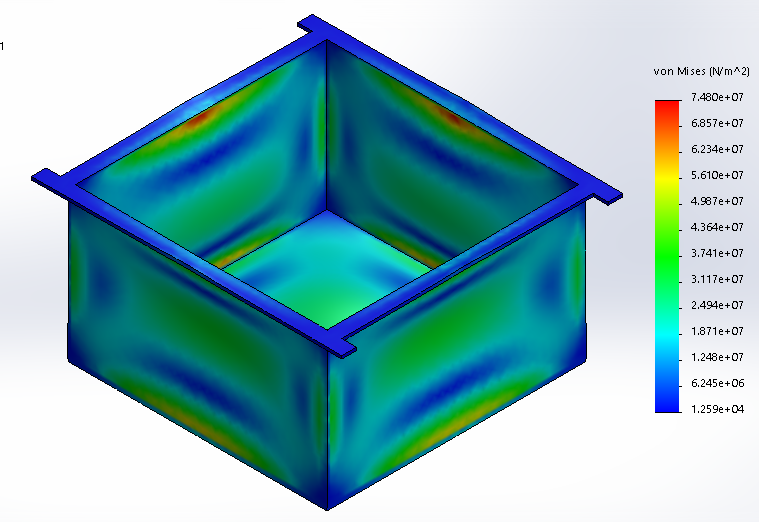}
\endminipage\hfill
\minipage{0.499\textwidth}
  \includegraphics[width=0.95\linewidth,trim={0.0in 0.0in 0.0in 0.0in}]{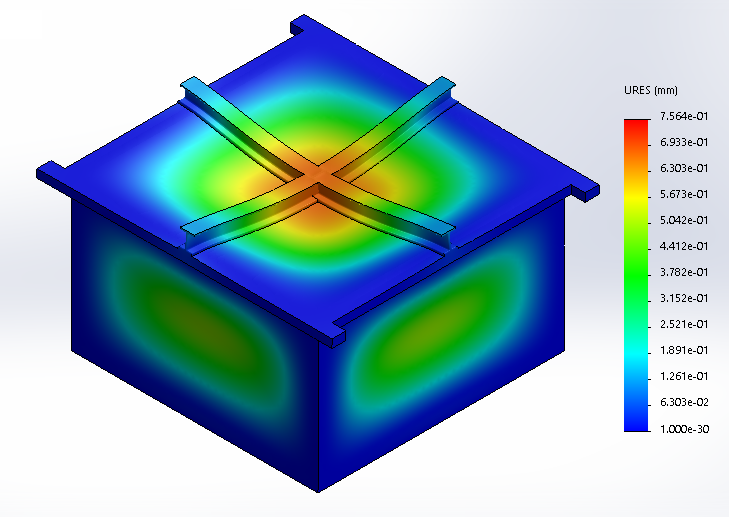}
\endminipage\hfill
\vspace{0.2in}
   \caption[example]
  {
Static simulation of the vacuum chamber under gravity and pressure.  The maximum stress (\textit{Left}) is located at the top and bottom of the sides and the center of the edge of the lid and base. The maximum stress is about a factor of 3 less than yield for both AISI 304 and 6061 T6. The maximum deflection (\textit{Right}) is less than a millimeter around the entire chamber.}
\label{fig:3}
\end{figure}

The vacuum chamber was designed so the maximum deflection in the system would be less than 1mm and the maximum stress would be less than a factor of 2 below yield.  The wall thicknesses (1” base, 5/8” sides) were chosen to meet this requirement. 

At high vacuum and located at sea level (1 atmosphere of pressure), the force on the lid and base of the chamber is 27,500 lbs and on the sides is 16,000 lbs. Figure \ref{fig:3} shows the stress and deflection due to this applied force as simulated in SolidWorks .  The maximum stress (74.8 MPa) is located along the top and bottom edges of the inside of the chamber and is about a factor of 3 less than yield (yield strength us 207 MPa for AISI 304 and 275 MPa for 6061 T6).  The maximum deflection is located at the center of the lid and base of the chamber and is only 0.8mm

A large rough and turbo pump and the NW-160 size port were chosen for their longevity, as the larger the pump, the faster the pump down speed, and the less strain on the vacuum over time. The ultimate pressure of the chamber will be below $\rm10^{-5}$ Torr with a projected pump down time of about an hour from atmosphere.

\subsection{Vibration Isolation}

The location of one of the vacuum systems will be in a laboratory building at UC San Diego. This building has large vibrations due to the HV/AC system, engineering labs performing vibration tests, and high foot and road traffic. To decrease vibration in the experiment, the vacuum chamber will be mounted to an 8" optical table on active pneumatic isolators.  

The optical table will provide high compliance between the light source and vacuum chamber, while the pneumatic isolators will provide vibration suppression. Based on measurements taken in the lab with an accelerometer comparing the vibrations between an active pneumatic isolated table and a regular table, the vibrations were suppressed by over an order of magnitude in terms of RMS displacement and peak to valley distance.

The vacuum pumps and cold head are additional sources of vibration and will be isolated from the chamber with bellows.  The cold head will extend into the chamber and be attached with copper strands with no hard contacts to avoid transmitting vibrations.  Each vacuum and the cold head will be mounted on vibration isolation pads to further reduced induced vibrations.

\section{Cold Shield}

\begin{figure}[t]
\minipage{0.45\textwidth}
  \includegraphics[width=0.95\linewidth,trim={0.0in 0.0in 0.0in 0.0in}]{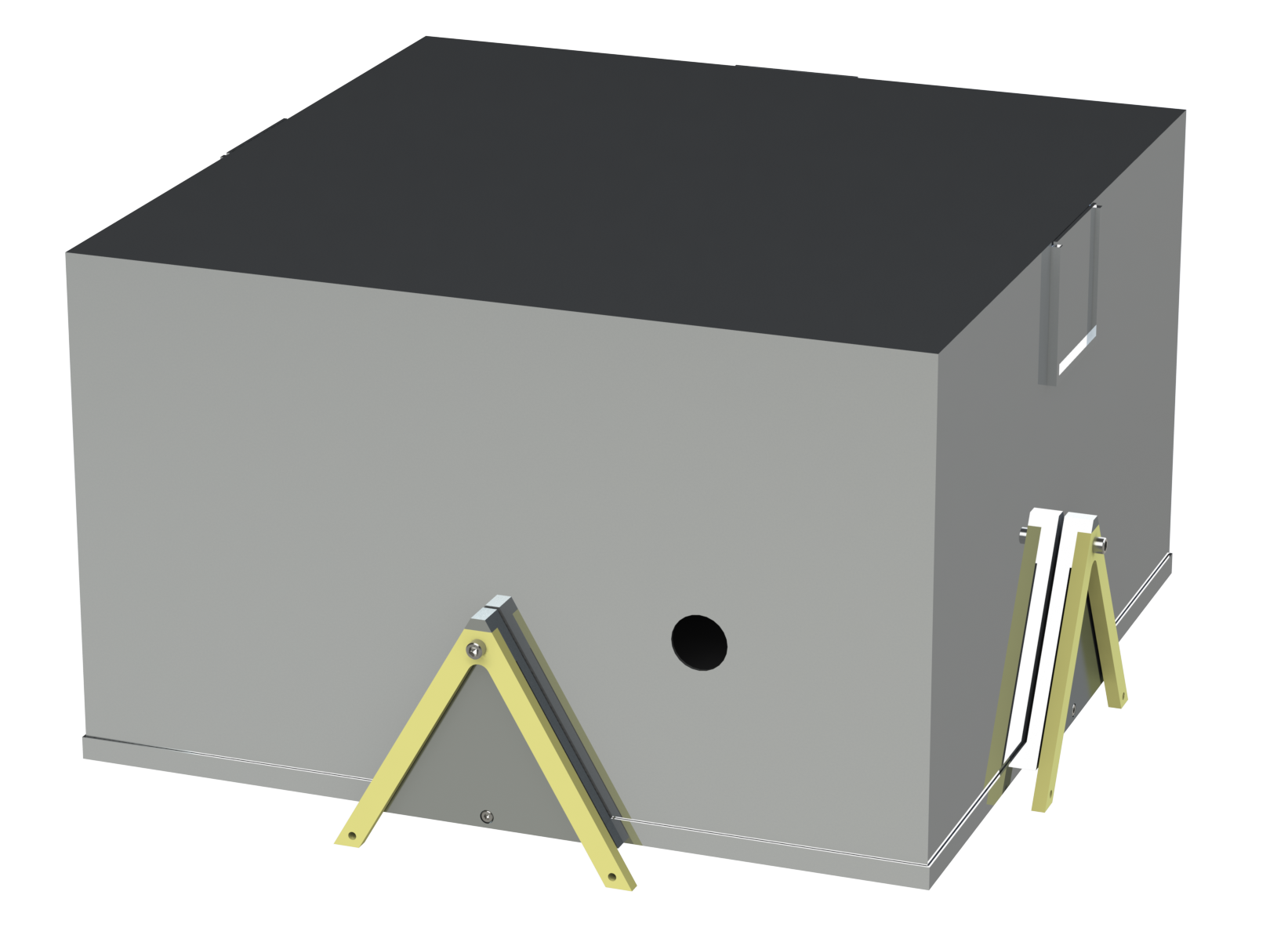}
\endminipage\hfill
\minipage{0.48\textwidth}
  \includegraphics[width=0.95\linewidth,trim={0.0in 0.0in 0.0in 0.0in}]{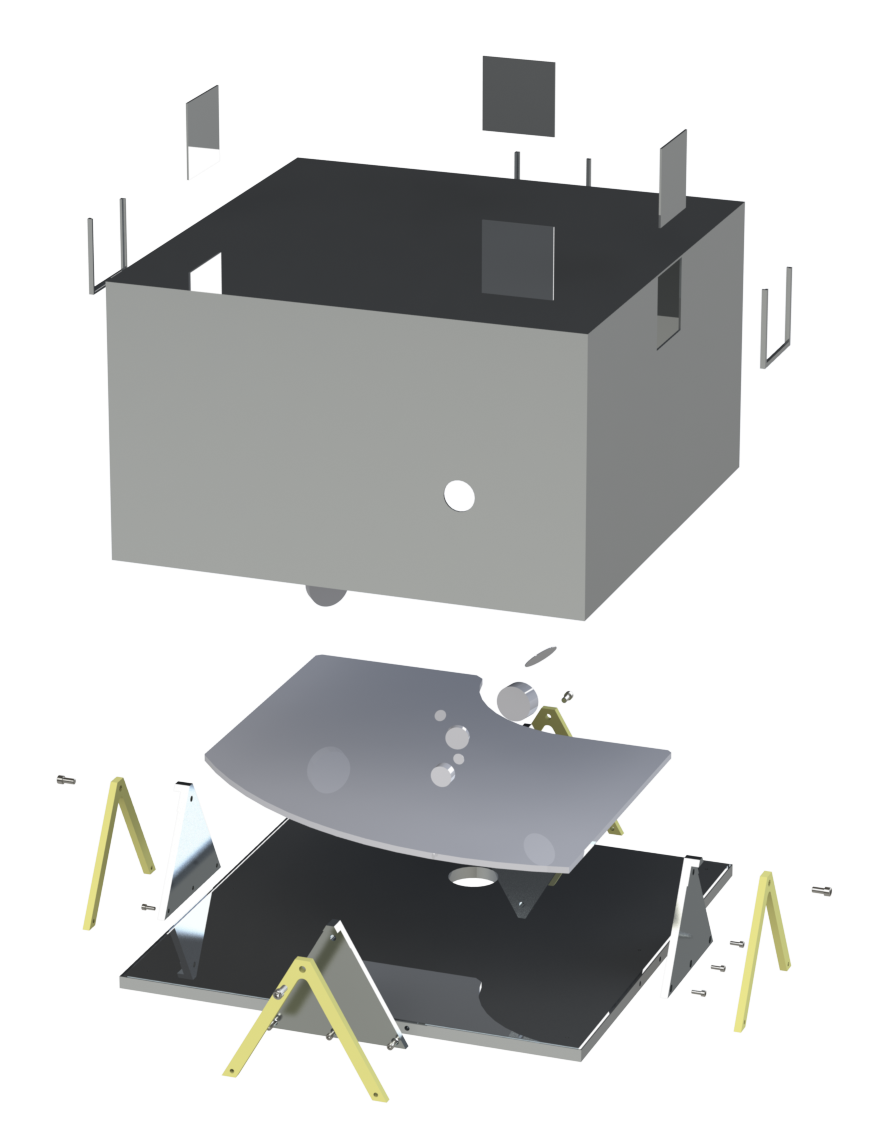}
\endminipage\hfill
\vspace{0.2in}
   \caption[example]
  {
Solid model and exploded view of the cold shield showing the optical plate, imaging optics, A-frames, and feed through plates.  The adapter plate, optical mounts, filter wheel, and pupil\cite{Cosens2020} are not included.}
\label{fig:4}
\end{figure}

The cold shield is made of 6061 T6 aluminum and is similar to the vacuum chamber in design (Fig. \ref{fig:4}).  The base is a thick 1” for rigidity and has through holes for mounting the adapter plate.  The walls and lid are non structural and are 4mm thick, yet weigh 90 lbs welded together and therefore require lifting points.  The walls lower into a groove in the base of the cold shield and are fastened with screws to improve thermal conduction.

There is an opening in one side concentric with the light source window on the chamber for the optical path, and another in the base concentric with the chamber cold head flange. There are 3 feed through plates that are 15cm$\times$15cm in size. These are cut into the side of the cold shield, and a channel part is welded on for the feed through plate to slide into.

The cold shield base has blind mounting holes on the outside for connecting to guides.  These guides attach to the A-frames at a higher point to allow the cold shield to sit low in the chamber while extending the length of the A-frame. This additional length helps to reduce the power conducted into the chamber as well as the stress induced from thermal contraction of the cold shield on the A-frame.

\subsection{Thermal Analysis}

To reduce the heat power conducted into the cold shield, long and thin G-10 A-frames are used.  G-10 has a low thermal conductivity of k=0.288 $\rm W \, m^{-1} \, K^{-1}$ \cite{2015arXiv150107100D}. The A-frames are bolted to guides connected to the cold shield and blocks connected to the base of the chamber. The equation for power conducted into the cold shield is: $$ P=\frac{kA(T_h-T_c)}{d}$$

\noindent Where $A$ is the cross-sectional area of the A-frame, $d$ is the length heat is conducted through the A-frame, $T_h$ is the temperature of the vacuum chamber, and $T_c$ is the temperature of the cold shield. 

For a quick estimate, we can model the thermal conduction through each leg of the A-frame as if it were a single beam and multiply by 8 to get the total power conducted into the cold shield. The cross-sectional area of one leg is $\rm A=3.75\times10^{-4}$ m$^2$. The minimum distance from the top bolt to the bottom is $\rm d=0.261$ m. With $T_h=293$ K and $T_c=55$ K, this gives us a power conducted through one leg of an A-frame as $P\approx0.1$ W and a total power conducted into the cold shield as $P_c\approx1$ W.

To reduce power radiated into the cold shield, the AISI 304 steel will be electropolished to reduce the emissivity and coated with layers of aluminzed mylar.  To reduce the power absorbed by the cold shield, it will

\begin{wrapfigure}{r}{0.5\textwidth}
\centering
\vspace{-0.in}
\includegraphics[width=0.5\textwidth]{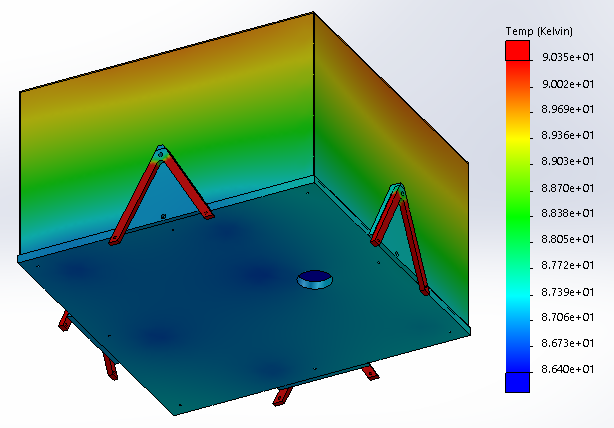}
\caption[Trackfits] 
 {
Thermal simulation run in SolidWorks with a 60W heat load corresponding to a cold head temperature of 55 K. The maximum temperature in the cold shield is 90 K located on the lid furthest from the cold head.}
\label{fig:5}
\end{wrapfigure}

\noindent be highly polished 6061 T6 aluminum. At 293 K, Mylar has an emissivity of 0.044 \cite{2015arXiv150107100D}. The equation for power radiated between two bodies is given as: $$P=\frac{\sigma(T_h^4-T_c^4)}{\frac{1}{A_hF_{hc}}+\frac{1-e_h}{A_he_h}+\frac{1-e_c}{A_ce_c}}$$.

\noindent where $\sigma=5.6703\times10^{-8}$ Wm$^{-2}$K$^{-4}$ is the Stefan-Boltzmann constant, $T_h=293$ K is the temperature of the vacuum chamber, $T_c=55$ K is the temperature of the cold shield, $A_h=5.217$ m$^2$ is the interior surface area of the vacuum chamber, $A_c=3.737$ m$^2$ is the exterior surface area of the cold shield, $e_h=0.044$ is the emissivity of the aluminized mylar layer around the vacuum chamber at $T_h$, $e_c=0.02$ is the emissivity of the 6061 T6 aluminum of the cold shield at $T_c$, and $F_{hc}\approx1$ is the view factor. This gives a power radiated into the cold shield as $P_r=23.9$ W.

For a more accurate estimate, we consider radiation from the window into the chamber which is modeled as an area with 100\% emissivity. The window is 60 mm in diameter giving an additional $P_w=1.2$ W into the cold shield. Another source of power is light coming through the gaps of the mylar layers which isn't modeled in this estimate.

Since the chamber will be operating with a vacuum level below $10^{-5}$ Torr, convection is negligible.  The total heat load on the cold shield is: $$P_{tot}=P_{r}+P_{c}=23.9+1.2+1=26.1 \text{W}$$ before accounting for power from electronics. This power corresponds to a cold head temperature of 40 K and gives a maximum temperature in the cold shield below 77 K as simulated in SolidWorks.  For an upper limit, we run a simulation case with about double this heat load at 60 W, which yields a cold head temperature of 55 K. Figure \ref{fig:5} shows the results from that case, with a maximum temperature of 90K at the opposite end of the cold shield. These simulations show that minimal radiation will be affecting the infrared detector system.

\subsection{Thermal Expansion}

To approximate the deformation of the cold shield, we use the linear thermal expansion equation: $$\delta=\alpha L\Delta T$$ where $\rm \alpha=2.3\times10^{-5} \, K^{-1}$ is the coefficient of thermal expansion for aluminum, $L$ is the length of the object, and $\Delta T$ is the temperature change. The cold shield base measures 0.940m$\times$0.940m.  After reaching an ultimate temperature of 55 K, its new size will be 0.935m$\times$0.935m. The displacement of the A-frame where it attaches to the cold-shield will be $\sim$2.5mm inwards on all sides. To make sure the A-frames can withstand this bend, we calculate the minimum length the A-frames need to be to survive using the equation:

$$\rm L>\sqrt{\frac{3\delta E w}{2\sigma_{max}}}$$

\noindent where, for G-10, $E=16.5$ GPa, $\sigma_{max}=262$ MPa \cite{2015arXiv150107100D}. The width of the A-frame is $w=0.015$ m and $\delta=0.0025$ m as before.  This gives us a minimum length of $L>0.06$ m.  The current length of the A-frame far exceeds this minimum length and simulations show the A-frames will easily withstand the deflection.

\section{Cart}

\begin{figure}[b]
   \begin{center}
   \begin{tabular}{c}
   \includegraphics[width=0.7\textwidth,angle=0,trim={0in 0in 0in 0in},clip]{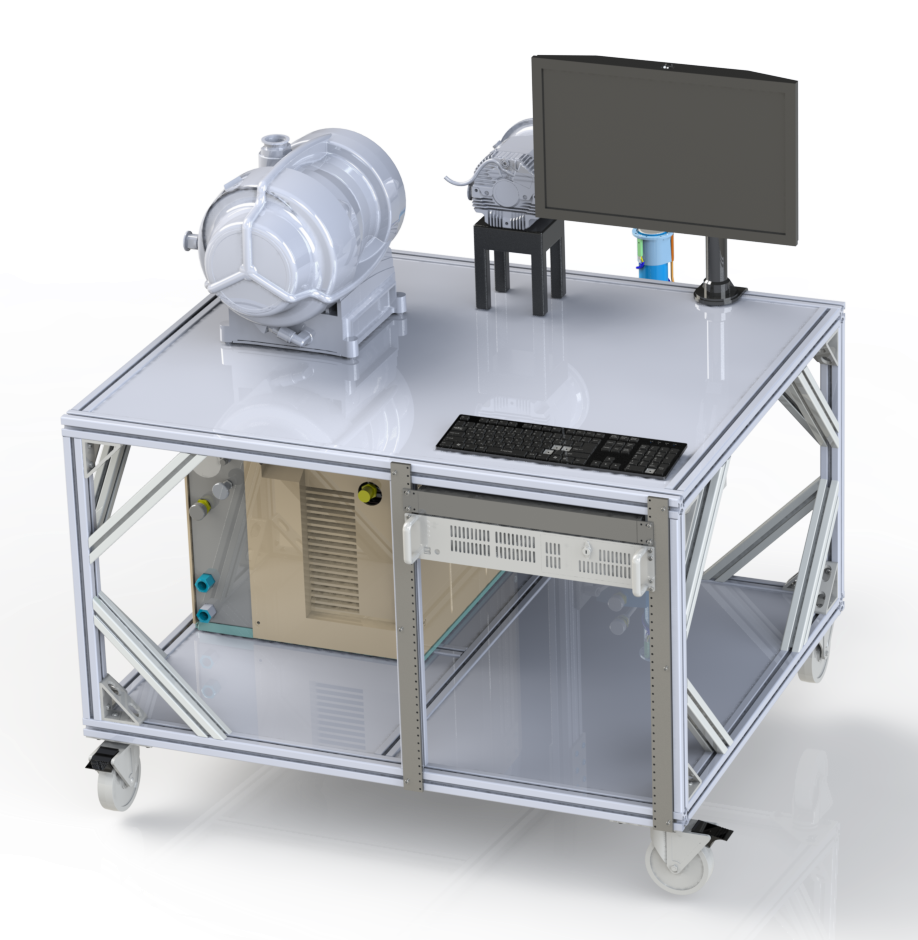}
   \end{tabular}
   \end{center}
   \caption[example]
  {
SolidWorks rendering of the cart housing the vacuum system, cryogenic system, and electronics. The cold head is held on a cantilever arm that sticks out behind the cart. The cart is constructed from standard 8020 components and is designed to be maneuverable for mounting to the vacuum chamber in multiple orientations.}
\label{fig:6}
\end{figure}

The cart servicing the test chamber will be made from 8020 extruded aluminum and other standard 8020 components (Fig. \ref{fig:6}). This cart will house the computer, rough pump and turbo pump on top, and the interior of the cart will house an electronics rack and compressor. The cold head will be extended via a two-dimensional cantilever arm to be able to connect to the bottom of the test chamber. This also allows the cold head to be disconnected and the cart easily removed when not in use. 

The cart is 99cm$\times$122cm$\times$79cm in size, with the cantilever arm extending 79cm from the cart. The cart is on triangular casters with stopping tabs, allowing it to be locked into place during operation. The rough pump, turbo pump, and cold head will be mounted on vibration isolation pads.

The cart is made from multiple different materials: the frame of the cart, as well as the braces,  are made from 8020 extruded aluminum, while the corners are fastened to the frame by a standard bolt assembly. The tabletops are made from PVC 0.007 Plasticized Plastic to resist against scratches over time. The cart rests on four triangular top casters with brakes and are attached to the frame via a standard bolt assembly. The cantilever arm extends out horizontally on a double thickness extruded aluminum, so as to provide stronger lateral support and prevent deflection. There is also a two-dimensional linear railing system on the cantilever arm so the cold head can be adjusted in vertical height and horizontal distance. The cantilever arm is attached to the main cart via side brackets.

\section{Summary and Future Work}

Liger is projected to be installed and operational by 2026-2027 \cite{2019BAAS}, capitalizing on the improvements to Keck AO from KAPA. It will provide higher spectral resolving power, larger fields of view, and a wider wavelength coverage than any existing IFS. We presented the design of the cryogenic test chamber that will be used to characterize the imaging camera and slicer component of the final Liger assembly.

This test chamber will operate at high vacuum and cryogenic temperatures to perform the necessary tests. It will also be vibration isolated due to the instrument's sensitivity and the vibrations measured in the lab.  It is serviced by a mobile cart that houses the vacuum system, cryogenic system, and electronics that interface with the experiment.

Future work on the experimental setup includes designing the adapter plate between the cold shield base and the optical plate, and designing the mounting system for the optics, detector, filter wheel, pupil wheel, and pupil viewing camera \cite{Cosens2020}. Manufacturing work on the chamber is expected to begin in 2021. The Liger opto-mechanical test dewar will provide an efficient means for characterizing the imager and slicer components while being adaptable for future cryogenic experiments.

\acknowledgments      
This research program was supported by the Heising-Simons Foundation.

\bibliography{report}
\bibliographystyle{spiebib}

\end{document}